\documentclass[prc,showpacs,superscriptaddress,twocolumn]{revtex4}
\usepackage{amsfonts}
\usepackage{amsbsy}
\usepackage{mathrsfs}
\usepackage{graphicx}
\usepackage{amsmath}
\def\be{\begin{equation}}
\def\ee{\end{equation}}
\def\bea{\begin{eqnarray}}
\def\eea{\end{eqnarray}}

\def\ba{\begin{array}}
\def\ea{\end{array}}

\def\lsim{\mathrel{\rlap{
\lower4pt\hbox{\hskip-3pt$\sim$}}
    \raise1pt\hbox{$<$}}}     
\def\gsim{\mathrel{\rlap{
\lower4pt\hbox{\hskip-3pt$\sim$}}
    \raise1pt\hbox{$>$}}}     

\begin{document}
\bibliographystyle{apsrev}
\title{TRANSVERSE $\Lambda^0$ POLARIZATION IN INCLUSIVE QUASI-REAL PHOTOPRODUCTION:\\
QUARK SCATTERING MODEL}

\author{\firstname{I.}~\surname{Alikhanov}}

\email{ialspbu@mail.ru}
\affiliation{%
Saint Petersburg State University, Saint Petersburg, 198904,
Russia
}%
\author{\firstname{O.}~\surname{Grebenyuk}}
\email{olegreb@pcfarm.pnpi.spb.ru}
\affiliation{%
Petersburg Nuclear Physics Institute, Gatchina, 188350, Russia
}%


\begin{abstract}
The transverse polarization of $\Lambda^0$ hyperons produced in
the inclusive $ep$ reaction at the 27.6 GeV beam energy is assumed
to appear mostly via scattering of the strange quark in a color
field. Results of application of such an idea to the preliminary
data of HERMES are presented. Contributions of $\Sigma^0$, $\Xi$,
and $\Sigma^*$ resonances to the polarization are taken into
account.
\end{abstract}

\pacs{13.60.-r, 13.88.+e \\ \vskip 0.1cm  (to appear in Physics of
Atomic Nuclei) }

\maketitle

\section{Introduction\label{introd}}
Polarization of $\Lambda^0$ hyperons has  attracted a lot of
experimental as well as theoretical activity almost since the very
moment of its discovery. Investigations of the phenomenon received
a specially great impetus in 1976 due to the striking experimental
results obtained at FERMILAB, where the hyperons produced in $pN$
collisions at a 300 GeV proton beam energy were highly polarized
\cite{fermilab}. The polarization was transverse and negative,
directed opposite to the unit vector
$\bold{n}\propto[\bold{p}_{b}\times{\bold{p}_\Lambda}]$, where
$\bold{p}_b$ and $\bold{p}_\Lambda$ are the beam and hyperon
momenta, respectively.

Only this direction is allowed by the parity conservation in
strong interactions provided the incident particles are
unpolarized. The results turned out to be in disagreement with the
expected negligible polarization in high energy processes as the
helicity is conserved in the limit of massless quarks (hereafter,
let us imply under polarization just transverse one).

The polarization has also been observed in a variety of other
hadron-hadron reactions at different kinematic regimes. Its
features qualitatively coincide in almost all the reactions, for
instance, being insensitive to the incident particle energy,
exhibiting the roughly linear growth by magnitude with the hyperon
transverse momentum $p_T$ and being negative. The only known
exception is the $K^-p$ process, where the polarization sign has
been found to be positive.

The $\Lambda^0$ wave function facilitates to some extent the
theoretical study. The SU(6) symmetry requires the spin-flavor
part of the wave function to be combined of $ud$ diquark in a
singlet spin state and strange quark of spin 1/2, or formally
$|\Lambda\rangle_{1/2}=|ud\rangle_0|s\rangle_{1/2}$, where the
subscriptions denote the spin states. Thus, one might entirely
attribute the $\Lambda^0$ polarization to its valence strange
quark.

Certainly, there have been proposed many approaches attempting to
account for the results (see, for example, reviews
\cite{review1,review2,review3} and the references therein).
However there is still no model, which would be able to describe
the complete set of the available measurements.

According to the empirical rules proposed by DeGrand and
Miettinen, the polarization sign depends on whether the $s$ quark
is accelerated (increases its energy) or decelerated (decreases
its energy) in the $\Lambda^0$ formation process \cite{degrand}.
To illustrate, there are no valence $s$ quarks in the initial
state of the $pp$ reaction so that they come from the quark sea to
form the final $\Lambda^0$. But the sea quarks predominantly
populate small $x$-states and consequently increase their average
energy coming in the valence content of $\Lambda^0$ ($x$ is
Bjorken variable). Here the polarization is negative. Contrary,
incident pseudoscalar kaons of the $K^-p$ reaction already contain
valence strange quarks, which are mostly decelerated in the
hadronization process. In this case, the sign is positive. Similar
ideas were implemented in flux-tube models with orbital angular
momentum \cite{anderson}.

The polarization in photoproduction has been investigated, for
example, in experiments on high energy $\gamma N$ scattering
performed at SLAC \cite{slac_gamma} and CERN \cite{cern_gamma}.
However, statistical accuracy of the experiments is indecisive and
would hardly enable one to conclude on the magnitude or on the
sign of the polarization.

In light of the scarce statistics for the $\Lambda^0$
photoproduction, the HERMES experiments on the 27.6 GeV positron
beam scattering off the nucleon target acquire a particular status
providing a good opportunity for observation of the polarization
in electroproduction. The collaboration has preliminary measured
nonzero positive transverse polarization \cite{Greb}, when most of
the intermediate photons are quite close to the mass shell, i.e.
$Q^2=-(p_{ei}-p_{ef})^2\approx 0$ GeV$^2$, where $p_{ei,f}$ are
the 4-momenta of the initial and scattered positrons, respectively
(quasi-real photoproduction).

Experimental properties of the polarization at HERMES turned out
to be reminiscent of those in the $K^-p$ reaction \cite{kexp},
which have been successfully described by a model assuming the
polarization to appear mostly via strange quark scattering in a
color field \cite{swed,gago}.

These arguments inspired us to apply the model to the  data
preliminary obtained by HERMES. In order to estimate the
contributions of $\Sigma^0,\Xi,$ and $\Sigma^*$ resonances into
account, we have generated the $ep$ process by PYTHIA 6.2 program
\cite{pythia62}.

\section{quark scattering model for $\Lambda^0$\label{model}}
It has been known that electrons, when scattering off nuclei, are
able to get polarization. Theoretically, it can be derived in QED
by considering a process of Dirac pointlike particle scattering
from static Coulomb potential provided next-to-leading order
amplitudes are taken into account \cite{feshb,dalitz}. The
corresponding formula reads
\begin{equation} \bold{P}=\frac{2  \alpha_{em}    m  p}{E^2} \ \frac{ \sin ^3
{\theta/2}  \ln[ \sin  {\theta/2}]}{\left[1-{p^2/E^2} \sin ^2
{\theta/2}\right] \cos {\theta/2}} \ \bold{n}, \label{eq:sz1}
\end{equation}
where $\bold{P}$ is the polarization vector, $E, p, m$ and
$\theta$ are the energy, momentum by magnitude, mass and
scattering angle of the electron, respectively, $\alpha_{em}$ is
the fine structure constant,
$\bold{n}\propto[\bold{p}_i\times\bold{p}_f]$, $\bold{p}_i$ and
$\bold{p}_f$ are the vectors of the electron momenta in the
initial ($i$) and final ($f$) states, respectively.

In \cite{swed}, Szwed proposed to consider the $\Lambda^0$
polarization as polarization of its valence strange quark using
Eq. (\ref{eq:sz1}). In other words, one should perform the
following interchanges in Eq. (\ref{eq:sz1}): electron
$\leftrightarrow$ quark, Coulomb potential $\leftrightarrow$ color
field ($\alpha_{em}$ $\leftrightarrow$ $C\alpha_s$), where
$\alpha_s$ is the strong coupling and $C$ is the color factor.

This approach has been applied to describe the polarization in the
$K^-p$ reaction and successfully reproduced  its main features at
$2C\alpha_s$=5.0 (while its theoretical value was found to be 2.5)
and the $s$ quark mass $m_s$=0.5 GeV \cite{gago}.

We have expressed the model in terms of the variable $\zeta$, the
HERMES data depend on
\begin{equation}
\zeta_{i(f)}=\frac{E_{i(f)}+p_{zi(f)}}{E_b+p_{zb}},
\label{eq:zeta}
\end{equation}
where the index  $b$ refers to the beam, the $z$ axis defines the
beam direction. Note that $\zeta$ is invariant under Lorentz
boosts.

According to recipes given in \cite{gago}, one should move to a
frame, where the magnitudes of the initial and final $s$ quark
momenta are the same (originally called $S$-frame). It is reached
by performing a Lorentz transformation along the proton momentum .
For this purpose one can write
\begin{equation}
(p_i\cdot p_f)=p^2(1-\cos\theta)+m_s^2,\label{eq:4vec_prod}
\end{equation}
\begin{equation}
p_{Tf}=p_T=p\sin\theta, \label{eq:p_t}
\end{equation}
where $p_{i,f}$ are the 4-momenta of the scattering quark,
$p_{Tf}$ is its transverse momentum in the center-of-mass frame of
the $K^-p$ reaction, while $p=\sqrt{E^2-m_s^2}$, $p_T$ and
$\theta$ refer to the $S$-frame.

On the other hand, using Eq. (\ref{eq:zeta}) leads to
\begin{multline}
(p_i\cdot p_f)-m_s^2=\frac{m_s^2}{2}
\frac{(\zeta_i-\zeta_f)^2}{\zeta_i\zeta_f}
\\
+\frac{1}{2}\left[p_{Ti}^2\frac{\zeta_f}{\zeta_i}
+p_{Tf}^2\frac{\zeta_i}{\zeta_f}\right]+(\bold{p_{T}}_i\cdot\bold{p_{T}}_f),
\label{eq:4vec_prod_zeta}
\end{multline}
where $(\bold{p_{T}}_i\cdot\bold{p_{T}}_f)$ denotes the ordinary
scalar product of the transverse momentum vectors.

Assuming that $p_{Ti}$=0, after some algebra, one can obtain from
Eqs. (\ref{eq:4vec_prod})-(\ref{eq:4vec_prod_zeta}) that
\begin{equation}
\cos \frac{\theta}{2} = \frac{ \xi  V_T^2}{(1-\xi)^2 +
V_T^2},\label{rel1}
\end{equation}
\begin{equation}
V=\frac{(1-\xi)^2
+V_T^2}{2\sqrt{\xi}\sqrt{(1-\xi)^2+(1-\xi)V_T^2}}, \label{rel2}
\end{equation}
where $V(V_T)$ and $\xi$ are defined by
\begin{equation}V_{(T)}=\frac{p_{(T)}}{m_s},\qquad
\xi=\frac{\zeta_f}{\zeta_i}.\label{ksi_vt} \end{equation}
Using relations (\ref{rel1}) and (\ref{rel2}), one can rewrite Eq.
(\ref{eq:sz1}) as
\begin{equation}
P(\xi, V_T)=\frac{2C\alpha_s V}{1+V^2 \cos^2 {\theta/2} } \ \frac{
\sin ^3 {\theta/2}  \ln [\sin  {\theta/2}]}{\cos {\theta/2} }\
sign(\xi-1), \label{basic formula}
\end{equation}
where the rules of DeGrand and Miettinen are expressed by the
factor $sign(\xi-1)$. It originates from the unit vector
$\textbf{n}$ in Eq. (\ref{eq:sz1}) as its direction depends on the
quark source (e.g., see \cite{swed}).  Actually, $\xi$ is the
longitudinal light cone momentum fraction of the initial quark
carried by the scattered one. Therefore, when the quark is
decelerated, i.e. $\xi<1$, the polarization defined by Eq.
(\ref{basic formula}) is positive, while it is negative at
$\xi>1$.  Note that there is a kinematic restriction in this
approach, imposed by Eq. (\ref{rel1}), since
$\cos{\theta/2}\leq1$. It forbids the variable $\xi$ to take
values in the following interval
\begin{equation}
1<\xi<1+V_{T}^2,\label{restrict}
\end{equation}
but the situation can be improved by introducing  continuous
$\zeta_{i(f)}$ distributions of the quarks, for example.

\begin{figure}
\includegraphics[angle=0,width=6cm,height=6cm]{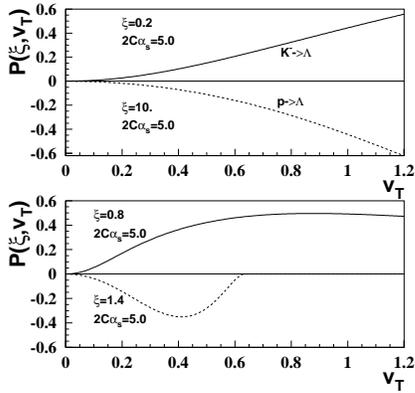}
\caption{Upper panel shows plots of Eq. (\ref{basic formula})
versus $V_T$ for $\xi=0.2$ at $2C\alpha_s$=5.0, which refers to
the $K^-p-$like event (solid line), and for  $\xi=10$ concerning
$pp-$like one (dashed line). Lower panel shows a plot of Eq.
(\ref{basic formula}) for $\xi=0.8$ (solid line) and $\xi=1.4$
(dashed line).} \label{Fig1}
\end{figure}

Plots of the polarization defined by Eq. (\ref{basic formula})
versus $V_T$ (equivalently $p_T$) for a few fixed values of $\xi$
at $2C\alpha_s=5.0$ are shown in Fig. \ref{Fig1}. The linear
growth, representative for hadron-hadron reactions, is seen well
on the upper panel for both the $K^-p-$like ($\xi=0.2$, solid
line) and $pp-$like ($\xi=10$, dashed line) events. On the lower
panel, we can see that for $\xi=0.8$ (solid line) the polarization
reaches a plateau, while for $\xi=1.4$ it grows taking maximal
value by magnitude at $V_T\approx0.4$ and then smoothly falls to
zero.

%

\begin{figure}
\includegraphics[angle=0,width=6cm,height=6cm]{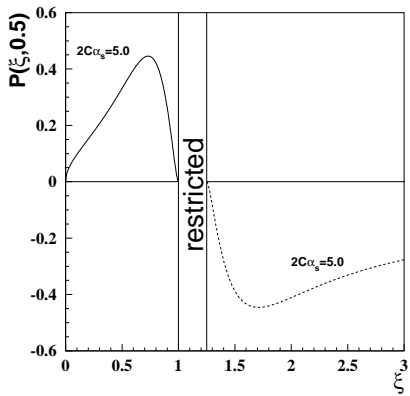}
\caption{Plot of Eq. (\ref{basic formula}) versus $\xi$  for
$V_T$=0.5 and $2C\alpha_s$=5.0. Both regions of the positive
(solid line) and negative (dashed line) polarization signs are
seen. The restricted area according to Eq. (\ref{restrict}) is
outlined.} \label{Fig2}
\end{figure}

A plot of the polarization versus $\xi$ for $V_T$=0.5 at
$2C\alpha_s=5.0$ is demonstrated in Fig. \ref{Fig2}. It is also
seen that for $\xi<1$ the polarization is positive and linearly
grows peaking at $\xi\approx0.75$, afterwards it steeply falls
down to zero. For $\xi>1.25$, it is negative, quickly growing by
magnitude up to $\xi\approx1.6$, afterwards decreasing very
slowly. The restricted area according to Eq. (\ref{restrict}) is
outlined as well.

\section{Calculations and results\label{exp events}}
The preliminary results of HERMES on the polarization have
qualitatively the same properties  as those of the $K^-p$ process
\cite{Greb}. It is positive and similarly depends on $p_T$ in both
the $K^-p$ and $ep$ reactions suggesting that closely related
underlying physics may be responsible for such a picture. These
similarities encouraged us to assume that quark degrees of freedom
of the positron beam might play significant role in the
polarization. Strange quarks may originate from the projectile
like the valence those in the incident kaons.

Thus, we have straightforwardly applied the model from \cite{gago}
to the quasi-real photoproduction. In order to do it, following
assumptions were made.

1) Since no information on the momentum of the initial quarks is
available in the experiment, it was assumed that the lepton beam
provides a collinear quarks with the $\zeta_i$ distribution, the
free parameters being $2C\alpha_s$ and $m_s$.

2) The final $s$ quark kinematic was determined as
\begin{eqnarray}
\zeta_f=\frac{m_s}{m_{\Lambda}} \zeta, \qquad
V_T=\frac{p_{T}}{m_{\Lambda}}, \label{eq:qqm}
\end{eqnarray}
here $\zeta$ and $p_{T}$ refer to the detected $\Lambda^0$'s.

It is evident that $\Lambda^0$'s detected in experiments are
produced not only via direct processes but may appear indirectly
as decay products of heavier hyperons, contributions of the latter
to the polarization are presumably considerable. Hence for more
adequate describing the process, one should consider such a
possibility.

To take possible contributions of $\Sigma^0$, $\Xi$ and $\Sigma^*$
resonances into account, we used events generated by the PYTHIA
6.2 package \cite{pythia62}. For this purpose, we partially
reproduced the HERMES acceptance   imposing the following limits
\begin{eqnarray}
p_{\Lambda}> 4.35\hskip 0.1cm \textrm{GeV}, \quad
\left|\frac{p_{x\Lambda}}{p_{z\Lambda}}\right|> 0.15, \quad
0.02<\left|\frac{p_{y\Lambda}}{p_{z\Lambda}}\right|< 0.14.
\label{eq:acceptance}
\end{eqnarray}
Additionally, to reflect in the calculations, at least
qualitatively, the phenomenological $\xi$ distribution, we
selected only events with $\xi>0.52$.

\begin{figure}
\includegraphics[angle=0,width=6cm,height=2cm]{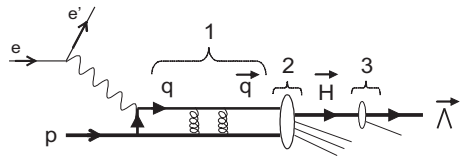}
\caption{Schematic illustration of the estimation of the indirect
process contributions to the $\Lambda^0$ polarization. 1) quarks
$q$ are polarized in the scattering. 2) the quarks combine into
the heavier resonances $H$ transferring them some fraction of the
initial polarization. 3) the resonances, thus polarized, decay
into $\Lambda^0$ transferring, in turn, to the final hyperon some
fraction of their polarization. An arrow sign over a letter
denotes the polarization.}\label{Fig3}
\end{figure}

The indirect process contributions to the  polarization were
estimated in three steps schematically illustrated in Fig.
\ref{Fig3}.

First, all the quarks ($q=u$, $d$, $s$) originated from the
intermediate photon were assumed to get polarization in the
scattering process. We calculated the corresponding polarizations
using Eqs. (\ref{rel1})-(\ref{basic formula}).

Second, $q\rightarrow H$, $H=\Sigma^0$, $\Xi$, $\Sigma^*$, i.e.,
having been polarized, the quarks combined into the heavier
resonances transferring them some fraction of the initial
polarization. The fractions are defined by the SU(6) fragmentation
spin transfer factors $t^F_{H,q}$ given in Tab.
\ref{tab:spintrans} \cite{liang}.

\begin{table}
\caption{Fragmentation spin transfer factors within the framework
of the SU(6) symmetry.} \label{tab:spintrans} \centering
\bigskip
\begin{tabular}{|c|c|c|c|c|c|c|}
\hline ~$q$~ & ~$\Lambda^0$~ & ~$\Sigma^0$~ & ~$\Xi^0$~ &
~$\Xi^-$~ & ~$\Sigma^*$~  \\
\hline
 $u$ & 0 &  2/3  &  ~0    &-1/3  & 5/9  \\
 $d$ & 0 &  2/3  & -1/3  &  ~0    & 5/9  \\
 $s$ & 1 & -1/3  &  2/3  &  2/3  & 5/9  \\
\hline
\end{tabular}
\end{table}

Third, $ H\rightarrow\Lambda^0$, i.e., the polarized resonances
decay into $\Lambda^0$ transferring, in turn, to the final hyperon
some fraction of their polarization. The fractions are defined by
the decay spin transfer factors $t^D_{\Lambda,H}$, which are taken
to be for $\Sigma^0\rightarrow\Lambda^0\gamma$,
$t^D_{\Lambda,\Sigma^0}=-1/3$, for $\Xi\rightarrow\Lambda\pi$,
$t^D_{\Lambda,\Xi}=0.91$ and for $\Sigma^*\rightarrow\Lambda\pi$,
$t^D_{\Lambda,\Sigma^*}=0.93$  \cite{gatto,gustafson}. Of course,
in general, there must be included another decay modes, such as
$\Sigma^*\rightarrow\Sigma^0\pi\rightarrow\Lambda\gamma\pi$,
$\Omega^-\rightarrow\Lambda{}K^-$,
$\Omega^-\rightarrow\Xi\pi\rightarrow\Lambda\pi\pi$, but they give
very small contribution, which was neglected.

Putting together the three points above, the total contribution of
the indirect processes can be formally written as
\begin{equation}
P=\sum\limits_{q,H}t^F_{H,q}t^D_{\Lambda,H}P_q,\label{eq:finpol}
\end{equation}
where $P_q$ is the polarization of a quark of flavor $q$ to be
determined by Eqs. (\ref{rel1})-(\ref{basic formula}). A detailed
explanation of similar calculations can be found in
\cite{gustafson}, for instance.

\begin{figure}
\includegraphics[angle=0,width=6cm,height=6cm]{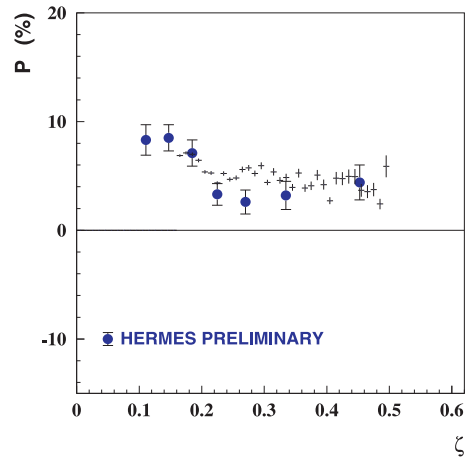}
\caption{Dependence of the $\Lambda^0$ polarization on $\zeta$
calculated including the contributions of $\Sigma^0$, $\Xi$ and
$\Sigma^*$ (scattered plot) in comparison with the preliminary
HERMES data (solid points). The data are taken from \cite{Greb}.}
\label{Fig4}
\end{figure}

Having chosen the model parameters as $2C\alpha_s=2.5$,
$m_u=m_d=0.33$ GeV and $m_s=0.5$ GeV, we carried out the
calculations. The polarization dependence on $\zeta$ calculated
including the contributions of $\Sigma^0$, $\Xi$ and $\Sigma^*$
(scattered plot) is shown in Fig. \ref{Fig4}. We can see a
reasonable reproduction of the HERMES data (solid points).

%

\begin{figure}
\includegraphics[angle=0,width=6cm,height=6cm]{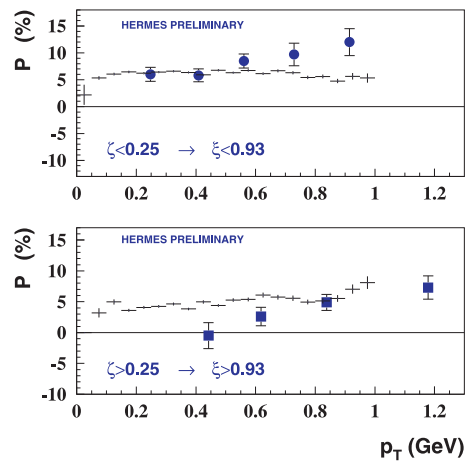}
\caption{Dependence of the $\Lambda^0$ polarization on $p_T$
calculated including the contributions of $\Sigma^0$, $\Xi$ and
$\Sigma^*$ (scattered plot) in comparison with the preliminary
HERMES data (solid points). There are two kinematic regions with
$\zeta<(>)0.25$ shown on the upper (lower) panel. The data are
taken from \cite{Greb}.} \label{Fig5}
\end{figure}

The $p_T$ dependence of the polarization calculated including the
contributions of $\Sigma^0$, $\Xi$, $\Sigma^*$ (scattered plot) in
comparison with the preliminary HERMES data (solid points) is
shown in Fig. \ref{Fig5}. Here the numerical results sufficiently
reproduce the experiment as well.

%
\section{Conclusion\label{conclusion}}
In this paper, we have calculated transverse $\Lambda^0$
polarization in the $ep$ reaction at the 27.6 GeV lepton beam
energy within an approach, which has been successfully applied to
the $K^-p$ reaction \cite{gago}. As the preliminary HERMES data
\cite{Greb} turned out to be qualitatively similar with those of
$K^-p$ \cite{kexp}, we assumed that the underlying polarization
mechanisms of both the $ep$ and $K^-p$ reactions could be similar
as well. Thus we supposed the $\Lambda^0$ polarization to
originate mostly in scattering of quarks coming from the positron
projectile. Recalling that the quasi-real photon exchange
dominated in the HERMES experiment, the said above might suggest
quark degrees of freedom of the photon to play considerable role
in the polarization process.

The quark scattering model \cite{swed,gago} has been expressed in
terms of the light cone variable $\zeta$ the HERMES data depend
on. On the other hand, a kinematic restriction for values of $\xi$
defined by Eq. (\ref{restrict}) was imposed. One can avoid the
problem introducing a continuous $\zeta_{i(f)}$ distribution as it
has done by PYTHIA. A sufficient reproduction of the HERMES data
has been reached, the contributions from $\Sigma^0$, $\Xi$ and
$\Sigma^*$ have been taken into account. For this purpose, we also
used the PYTHIA program. All our results  should be regarded only
as qualitative.

The largest difficulty of this model is the parameter $2C\alpha_s$
since the strong coupling is running, additionally it was
impossible to derive $\alpha_s$ from the HERMES data.

As a further development, another sources, such as gluon-gluon
fusion etc. \cite{dharma}, can be included and treated in the same
way.  It would be also interesting to compare our results with
those recently obtained in the framework of the quark
recombination approach \cite{kubo}, they show negative
polarization when $p_T$ varies from about 0.2 GeV up to about 0.6
GeV for $\zeta>$ 0.25.

When this work had been already completed, the HERMES
collaboration published new data on the $\Lambda^0$ polarization
\cite{new}, which are also in well qualitative agreement with the
calculations presented herein.


\end{document}